# The spectrum of a single photoionized cloud

Gary Ferland

*CITA, U of Toronto, and Physics, University of Kentucky*

**Abstract**.  The emission-line spectrum of a quasar is most likely emitted by an ensemble of photoionized clouds moving with a variety of velocities, with a range of densities and distances from the central object.  The state of the art in this field is to first compute the emission from a single cloud, and then prescribe a mix of clouds to reproduce line intensities, profiles, and reverberation lags.  Here I review the parameters that define emission from a single cloud and the outstanding problems in quasar emission-line analysis.

## 1 Introduction

This is a quick primer for those unfamiliar with the basis of photoionization modeling.  More complete discussions are those of Davidson & Netzer (1979), Osterbrock (1989), and Netzer (1990).  We discuss here only the properties of a single cloud.  Several of the following papers go on to the more realistic situation of an ensemble of clouds.

## 2 Photoionization equilibrium of a single cloud

The spectrum emitted by a low-density gas irradiated by an energetic continuum is really just a large-scale fluorescence problem.  The ionizing radiation's energy is initially converted into a photoelectron's kinetic energy, which is then shared with other particles in the cloud, and eventually degraded into the observed emission line or continuum spectrum.

The essential difficulty is that the gas is far from equilibrium – the concept of temperature has little meaning.  This means that the ionization and level populations are not given by Saha or Boltzmann statistics, but rather by the balance of microphysical processes. The only temperature that is ever used is the kinetic temperature of the electrons.  This is generally valid because elastic electron-electron collisions are very rapid, due to their low mass and high velocity, and insures that a Maxwellian velocity distribution is quickly set up.  As a result of these complications the spectrum is best understood by reference to large-scale numerical simulations.  Ferland et al. (1998) describe such calculations.

### 2.1 Ionization: photoionization-recombination balance

The balance between processes that ionize species and those that allow it to recombine determines the level of ionization of the gas.  Ionization processes include direct photoionization of outer shell electrons, photoionization of inner shells followed by ejection of Auger electrons, collisional ionization, and charge transfer ionization.  Recombination processes include radiative, dielectronic, and three-body recombination, and charge transfer recombination.

Usually the most important of these processes are valence shell photoionization and a combination of radiative and dielectronic recombination.   In photoionization problems the flux of ionizing photons striking the illuminated face of the cloud is a fundamental parameter.  This is given by

$$\Phi(H) \equiv \int_{\nu_1}^{\infty} \frac{4\pi J_\nu}{h\nu} d\nu \ \ [\text{photons cm}^{-2}\,\text{s}^{-1}] \tag{1}$$

where $J_\nu$ is the intensity. The photoionization rate [units s$^{-1}$] for a species with a typical photoionization cross section $<\sigma>$ [units cm$^2$] will be $\sim\ \Phi(H) <\sigma>$. In equilibrium this is equal to the recombination rate, $n_e\ \alpha(T)$, where $\alpha(T)$ is the recombination coefficient.  Both $\sigma$ and $\alpha$ are the results of extensive atomic physics calculations and tabulations are readily available (see Verner's *Atomic Data for Astrophysics* web site at http://www.pa.uky.edu/~verner/atom.html).

The ionization balance equation is

$$n_{atom} \langle \sigma \rangle \Phi(H) = n_e n_{ion} \alpha(T) \tag{2}$$

and the resulting level of ionization is

$$\frac{n_{ion}}{n_{atom}} = \frac{\Phi(H)}{n_e} \frac{\langle \sigma \rangle}{\alpha(T)} \approx U \frac{\langle \sigma \rangle}{c\,\alpha(T)} \tag{3}$$

where c is the speed of light.  The last equation introduces the ionization parameter $U$, defined as

$$U \equiv \frac{\Phi(H)}{n_H c} \tag{4}$$

$U$ is sometimes used to parameterize the radiation field instead of $\Phi(H)$ because it is directly proportional to the level of ionization and has the advantage of introducing homology relations among various models. The photon flux $\Phi(H)$ has the benefit of directly exposing the dependence on the separation between cloud and central source – this is directly measured by reverberation studies.  In practice either U or $\Phi(H)$ can be used as a parameter, and for a given density the level of ionization goes up with each.

## 2.2 Temperature: heating-cooling balance

Each photoelectron is ejected with a kinetic energy equal to $h\nu - IP$, where $h\nu$ is the energy of the incident photon and *IP* the ionization potential of the atom. The heating per unit volume due to photoionization will be

$$G = n_{atom} \int_{\nu_1}^{\infty} \sigma_\nu \frac{4\pi J_\nu}{h\nu} \{h\nu - IP\} d\nu \ \ [\text{erg cm}^{-3}\,\text{s}^{-1}] \tag{5}$$

where $n_{atom}$ is the atom's density.  This kinetic energy is rapidly shared with other free electrons to establish the electron temperature $T_e$.

Cooling is any process that removes kinetic energy from the free electrons. Usually the most important cooling process is a collision of an electron with an ion, resulting in the collisional excitation of a bound electron that then decays, emitting a photon.

This is clearly an intricate and non-linear problem.  The heating/cooling and resulting temperature cannot be determined until the ionization of the gas is known.  The collisional ionization and recombination rates depend on temperature, so the ionization cannot be determined until the electron temperature is known.  All of this must be solved as a function of depth into the cloud, taking into account the attenuation of the continuum by gas opacity and the effects of radiative transport.

## 2.3 Energy units, the Rydberg

The spectra shown in various papers in this book have either wavelength (with units microns, nanometers, or Ångstroms) or energy/frequency (units Hz, eV, keV) as the x-axis.  To make matters worse the natural unit of energy in photoionization problems is the Rydberg.  1 Rydberg is nearly equal to the ionization potential of hydrogen, 13.6 eV.  The formal definition is

$$R_\infty \equiv \frac{2\pi^2 m_e q_e^4}{ch^3} = 109737.315\,\text{cm}^{-1} \quad . \tag{1}$$

$R_\infty$ is the ionization potential of an infinite-mass nucleus with a charge of one. Hydrogen itself has a lower ionization potential due to the reduced mass of the nucleus, so the Rydberg constant for hydrogen itself is

$$R_H = 2.178728\times10^{-11}\text{erg} = 13.59842\text{eV} = 91.176340\text{nm} = 109677.576\text{cm}^{-1} \tag{2}$$

# 3 Photoionization model results

The goal of emission line analysis is to deduce the properties of the clouds that produce the observed spectrum.  This section describes the basic parameters that determine the conditions in a cloud and the resulting spectrum.  These parameters are the shape of the ionizing continuum, the gas' density $n_H$, the chemical composition, the column density $N_H$, and either $U$ or $\Phi(H)$.

In the following sections we will describe these parameters and explore their effects on results.  We consider deviations from a single fiducial model, a cloud with a density of $n_H = 10^{10}$ cm$^{-3}$, an ionization parameter of $U = 10^{-1.5}$, solar abundances, and a column density of $N_H = 10^{23}$ cm$^{-2}$.  This density and column density corresponds to a physical thickness of $10^{13}$ cm, or a bit under one Astronomical Unit.

All of the calculations use a simple power-law continuum, $f_\nu \propto \nu^\alpha$.  We really can't do any better since the best evidence is that we do not directly observe the same continuum as the clouds (see the discussion by Zheng in this volume, Netzer 1985, and also Korista et al. 1997).  We will focus on continua with slopes near the centroid of the observed range, $\alpha \sim -1.5$.  This power law continuum extends between 1.36 eV and 50 keV.  Outside this range the continuum is assumed to drop rapidly.  Continua are described as soft or hard, depending on the energy of a typical photon.  Harder continua have more energetic photons and a less negative $\alpha$.

The approach taken for many years was to try to use the observed spectrum to obtain the parameters of a typical BLR cloud by comparing this spectrum with detailed numerical simulations such as those presented below.  Today we know that clouds with a broad range of parameters exist within the emission line regions, so the notion of a single set of parameters has little meaning.

I will show how some line pairs change with model parameters.  These lines were selected based on the discussion in the meeting.  I tried to consistently show the same set of lines in every plot; if a line ratio is not plotted then it probably did not change with that particular parameter.

Finally, all of the calculations presented here were done with version 90.05 of the spectral simulation code Cloudy (Ferland et al. 1998; on the web at http://www.pa.uky.edu/~gary/cloudy). The programs used to generate these grids of results are available on this site.

### 3.1 A cloud's structure

Figure 1 shows the ionization and thermal structure of a cloud with our fiducial parameters.

The structure has a thin layer of highly ionized gas, $He^{+2}$, $C^{+3}$, $O^{+5}$, extending to a depth of $10^{11}$ cm, followed by region of more moderate ionization, $H^+$, $He^+$, $C^{+2}$. A $H^+$ - $H^o$ ionization front occurs at bit under $10^{12}$ cm. Unlike galactic nebulae, hydrogen does not become predominately neutral after the ionization front, but instead remains partially ionized. This is due to penetrating X-Rays heating the gas, and the resulting trapped Ly$\alpha$ creating enough population in the first excited level of H for photoionization by the Balmer continuum

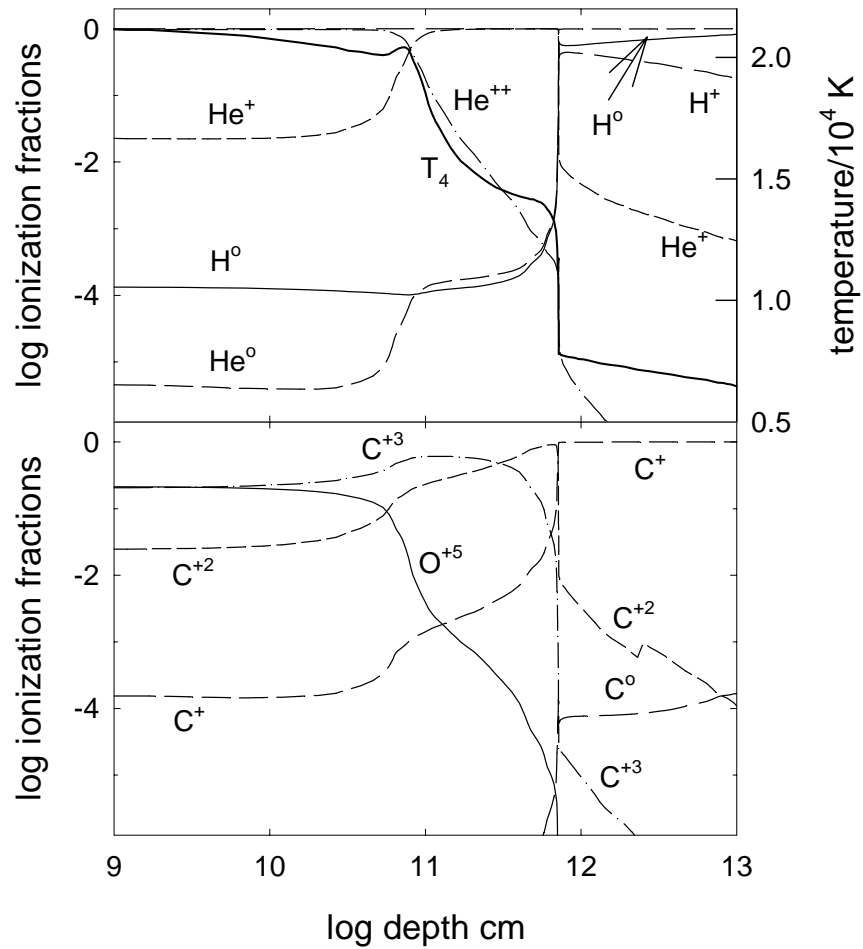

Figure 1 The ionization structure of a typical blr cloud. The density and ionization parameter were $10^{-1.5}$ and $10^{10}$ cm$^{-3}$, and an $\alpha$ = -1.5 power law ionizing continuum was assumed. The calculation stopped at a column density of $10^{23}$ cm$^{-2}$.

to become important.

The hydrogen ionization front occurs at the point where nearly all hydrogen-ionizing photons have been absorbed. It turns out that the column density to the hydrogen ionization front is a simple function of $U$. Since each photoionization is followed by a recombination, the number of recombinations that occur over the depth $D$ to the ionization front is $D\, n^2\, \alpha$. This must equal the number of ionizing photons that enter this column, $\Phi(H)$, so the depth to the ionization front is $D = \Phi(H)/n^2\, \alpha$. The gas column density is $N = Dn$, so the column density to the hydrogen ionization front is

$$N_{lfront} = Uc/\alpha \approx 10^{23} U \quad [\text{cm}^{-2}] \,. \tag{6}$$

If the cloud's total column density is large enough for a hydrogen ionization front to occur then the cloud is said to be *radiation bounded* or optically thick to ionizing photons. A low column density cloud, where the gas remains highly ionized, is said to be *matter bounded* or optically thin to ionizing photons.

The temperature is also shown in Figure 1. The gas is hottest in the He$^{+2}$ region, the temperature falls in He$^+$ region, and then falls still more to ~5000K in H$^o$ region.

### 3.2 The ionizing continuum

The shape of the ionizing continuum is both a fundamental parameter for the model cloud and of great interest because of its connection to the nature of the central object. We restrict ourselves here to the simple power-law continua described in section 3.1. Figure 2 shows the effects of changing the slope of the power law continuum while keeping all other parameters constant.

The equivalent width of Ly$\alpha$ grows larger as the continuum gets harder. Ly$\alpha$ is largely a recombination line — calculations show that its intensity remains generally within 0.5 dex of its pure recombination value. In this limit each hydrogen recombination produces one Ly$\alpha$ photon, so the intensity of Ly$\alpha$ is proportional to the number of ionizing photons that strike clouds. If the continuum is simply $f_\nu \propto \nu^{\alpha}$ then the ratio of the line to continuum is given by

$$\frac{I(Ly\alpha)}{\nu f_\nu(1216)} = \frac{EW(Ly\alpha)}{1216} \approx \frac{\Omega}{4\pi} \frac{h\nu_{3/4} \int_{\nu_1}^{\infty} \frac{\nu^{\alpha}}{h\nu} d\nu}{\nu_{3/4}^{\alpha+1}} \approx -\frac{\Omega}{4\pi}(3/4)^{-\alpha}\, \alpha^{-1} \tag{7}$$

where $\nu_{3/4}$ is the frequency of Ly$\alpha$ (3/4 of a Rydberg) and the integral is over all ionizing energies. The results shown in Figure 2 lie close to this simple estimate.

The covering factor $\Omega/4\pi$ introduced in eq 7 is the fraction of the photons emitted by the central object that actually strike clouds. We know that this must be significantly less than unity since quasars have unobscured lines of sight to the continuum source. Orientation effects are discussed extensively in other papers, and it is thought that objects not viewed from this preferred direction are highly obscured.

Quasars have typical Ly$\alpha$ equivalent widths of several hundred Ångstroms. From Figure 2 we see that, for a typical $\alpha$ of –1.5, the covering factor must be greater than 1/3 if photoionization is to explain the energetics. Several papers in this conference allude to either this constraint, or a closely related one that uses the equivalent width of HeII 1640. Finally, in the nebular literature the line to continuum ratio is referred to as the Zanstra method of determining continuum shapes (or equivalently, the stellar temperature).

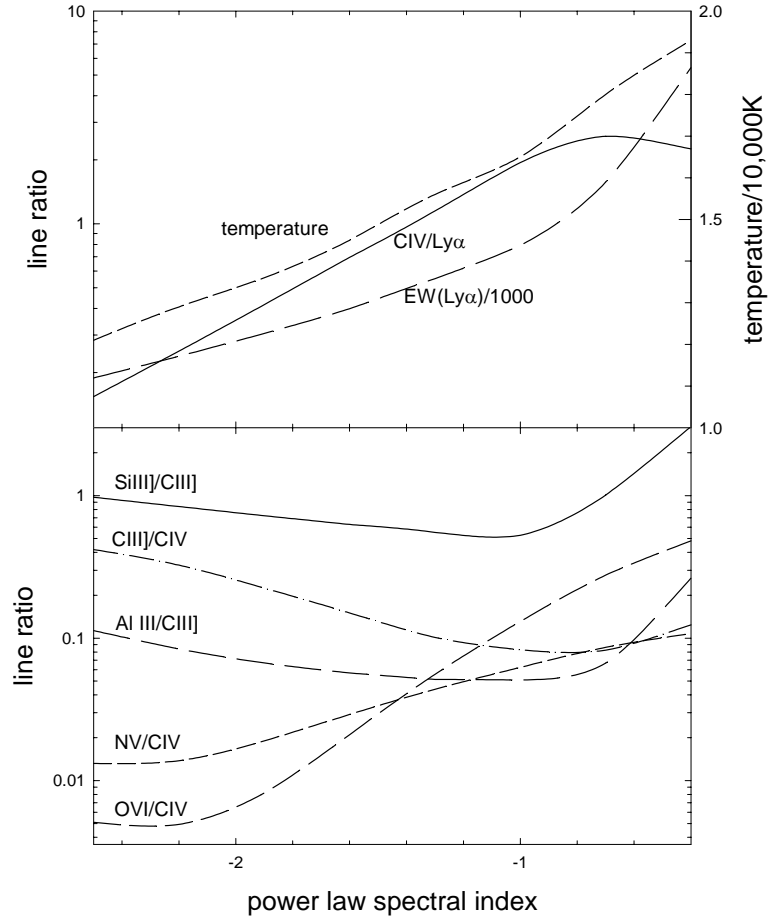

Figure 2 The form of the power-law continuum between 1 micron and 50 kev was varied over the indicated range. The plot shows several line ratios, the equivalent width of Lyα, and the temperature averaged over the O++ zone.

Figure 2 shows that the O++-weighted mean temperature and CIV/Lyα ratio both increase with increasingly harder continua. This is because the typical photoelectron energy increases as the continuum grows harder (equation 5) so there is more heating per photoionization. CIV 1549 is one of the strongest coolants in a typical cloud. The ratio CIV/Lyα is essentially the cooling per recombination, equal to the heating per photoionization, and so to the mean energy of ionizing photons. This line ratio is closely related to the Stoy method of determining stellar temperatures in the nebular literature.

We can use line-intensity ratios to deduce the shape of the continuum from the emission-line spectrum. The basic idea is that the intensity of an H or He recombination line is proportional to the number of ionizing photons with energies capable of producing that ion. Then the intensity of a HeII recombination line relative to Lyα is proportional to the ratio of the number of photons with energies greater than 4 Ryd to the number greater than 1 Ryd.

Mathews & Ferland (1987) used this approach to find a continuum that peaked somewhere near 4 Ryd. This shape *does not* agree with the continuum directly observed in z~1.5 objects (Zheng et al. 1997). Korista, Baldwin, & Ferland (1997) pointed out that the observed continuum is not luminous enough to account for the energy in the high-ionization lines. The implication is that the emission-line clouds do not see the same continuum we do, a conclusion previously reached by Netzer (1985). This is not surprising since the geometry is thought to have a cylindrical symmetry, and there is a growing consensus that we preferentially view AGN from nearly pole-on directions.

### 3.3 The infrared continuum

The infrared continuum is critical for dense clouds because of the importance of free-free or bremsstrahlung heating (Ferland et al. 1992). Free-free heating is the absorption of a photon by an electron – ion pair, with the electron gaining most of the photon's energy. The cross section is just the bremsstrahlung opacity, which is proportional to $\nu^{-2} n^2$, with $\nu$ the photon's frequency and $n$ the gas density. From this we see that free-free absorption is most important for dense clouds, and for the lowest frequencies.

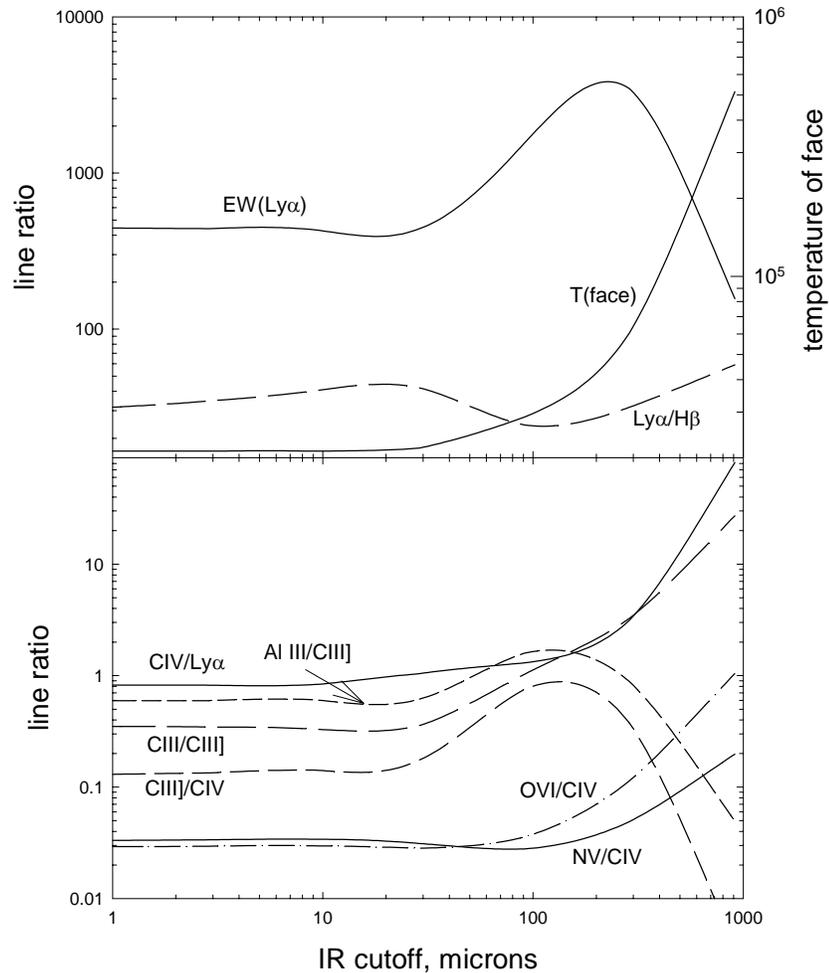

Figure 3 Dependence of the spectrum on the infrared cutoff.

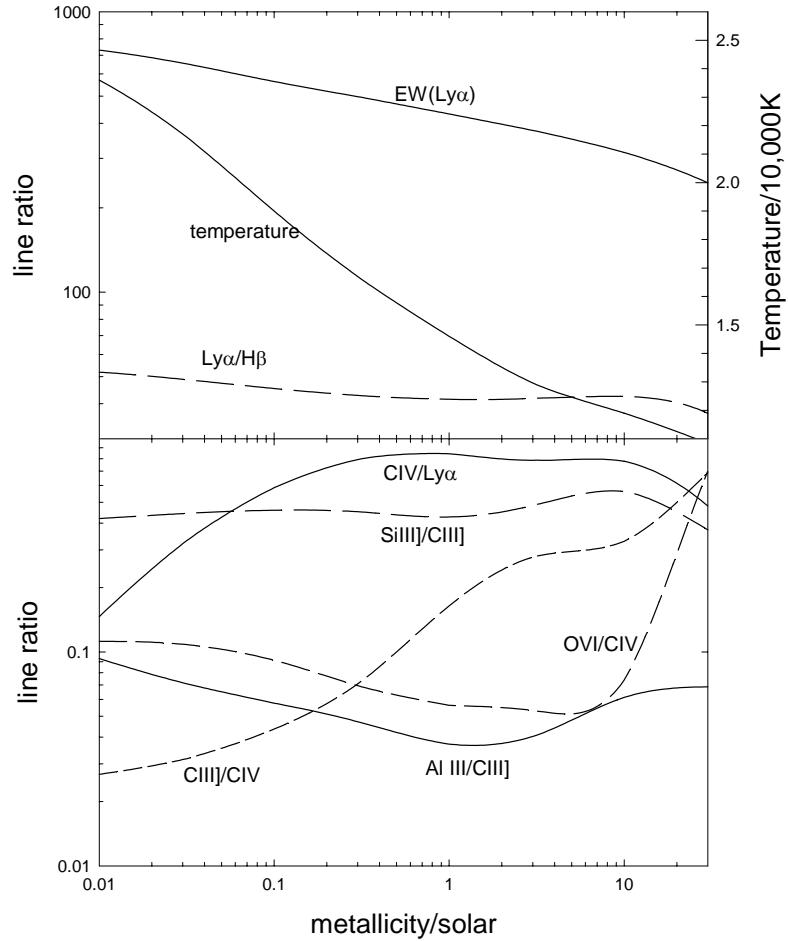

Figure 4 Line ratios and mean temperature as a function of the metals scaling factor. The $\alpha = -1.5$ power law continuum, with a density and ionization parameter of $10^{10}$ cm$^{-3}$ and $-1.5$ were used.

Figure 3 shows a series of calculations in which the IR break of our fiducial $\alpha = -1.5$ power-law continuum was varied. Other cloud parameters were unchanged. (The continuum used in all other calculations had an IR break at 0.912 micron to minimize the importance of free-free heating.) The figure shows several line ratios and the temperature of the illuminated face of the cloud.

For this set of cloud parameters free-free heating becomes dominant when the UV power law extends to ~30 microns. For denser clouds the effects become more dramatic for much shorter wavelength cutoffs. For instance at a density of $10^{12}$ cm$^{-3}$ free-free heating becomes important when the cutoff is under 1 micron. This is an additional uncertainly since it is not clear exactly where the observed IR continuum forms. There are dust and starlight contributions that must originate well beyond the BLR. But there are also cases where the IR is a component of the central non-thermal continuum (see the re-view by Wilkes in this book) and so should be included in the radiation field striking the clouds.

There has been surprisingly little exploration of the effects of the IR continua upon BLR clouds, despite the fact that this is *more* important than the observed hard x-rays. For instance, could differences in the near to far IR continuum, as seen by the BLR, be a contributor to differences in spectra of radio loud and radio quiet objects?

### 3.4 Composition

The emission-line spectrum is surprisingly insensitive to the gas composition. Figure 4 shows a series of calculations in which the abundances of all elements heavier than helium were varied by the scale factor shown as the x-axis. Unity represents solar abundances. The CIV 1549/Lyα intensity ratio changes by only a factor of four while the C/H ratio changes by 3.5 dex. This reflects energy conservation – CIV is a strong coolant, and the total cooling must balance heating.  As the metals scaling factor goes down the cooling efficiency of the gas does too, so the $O^{++}$-weighted mean temperature (also plotted) goes up.  This *thermostat effect* ensures that the overall spectrum is largely unchanged despite global changes in the chemical enrichment.

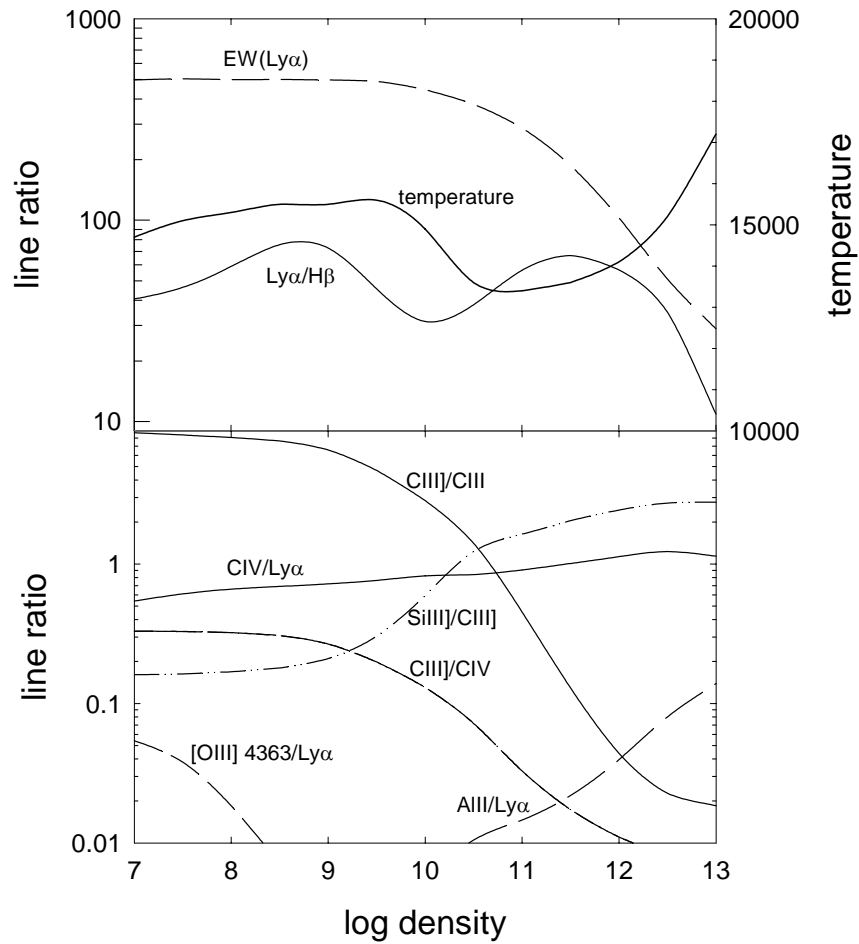

Figure 5 Spectrum as a function of density. The calculations assumed an α=-1.5 power law, solar abundances, and stopped at the hydrogen ionization front. The lines plotted are Lyα 1216, Hβ 4861, [OIII] 4363, Al III 1857, CIV 1549, CIII] 1909, and SiIII] 1892.

Increasing the abundances of the heavy elements also raises the gas opacity.  At high abundances elements other than H and He absorb ionizing radiation and the intensity of Lyα relative to the continuum goes down.  We have argued that this is a partial contributor to the Baldwin effect (Korista et al. 1998).

The first studies of quasar spectra found that the abundances are broadly consistent with solar (see Davidson and Netzer 1979), a conclusion that remains valid today.   Although the thermostat effect prevents us from measuring absolute abundances relative to hydrogen, it is possible to measure relative abundances, such as N/C or N/O.  Beginning with Shields (1976), most studies try to measure the metallicity by measuring the abundance of nitrogen relative to C and O and then relating this to overall enrichment scenarios.  Our recent work suggests that there is a luminosity – metallicity correlation, with the most luminous objects having a metallicity Z that is 5 – 10 times solar.  This is discussed extensively in Fred Hamann's chapter in this book, as well as in Hamann & Ferland (1999).

### 3.5 Density

The *emissivity* of a line, the energy released per unit volume and time, will be a function of the gas density that depends on the detailed atomic physics. The balance equation for a two-level atom can be written as a balance of two rates [with unit $cm^{-3} s^{-1}$]:

*lower to upper = upper to lower*

$$n_l \, q_{l,u} n_e = n_u \left( A_{u,l} \beta + q_{u,l} n_e \right) \tag{8}$$

where β is the escape probability and $q$ is the rate coefficient for collisional processes ($cm^3 s^{-1}$).  Note that $\beta \sim \tau^{-1}$, where τ is the line's optical depth (Elitzur 1992).  The emissivity will be

$$\varepsilon_{coll} = n_u A_{u,l} \beta \, h\nu = n_l A_{u,l} \beta \, h\nu \frac{q_{l,u} n_e}{\left( A_{u,l} \beta + q_{u,l} n_e \right)} \; [ergs \; cm^{-3} \, s^{-1}]. \tag{9}$$

This has a powerful temperature dependence since the rate coefficient $q_{l,u}$ varies as exp(-hν/kT).

The critical density $n_{crit}$ for a line is the density where the two terms in the denominator of equation 9 are equal:

$$n_{crit} = A_{u,l} \beta \big/ q_{u,l} \approx A_{u,l} \big/ q_{u,l} \tau \, . \tag{10}$$

Physically $n_{crit}$ is the density where the upper level is as likely to be collisionally de-excited as to emit a photon.  Forbidden lines have $n_{crit} \sim 10^3 - 10^6 \, cm^{-3}$, intercombination lines like CIII] 1909 have $n_{crit} \sim 10^9 - 10^{11} \, cm^{-3}$, and permitted lines $n_{crit} \sim 10^{14} - 10^{16} \, cm^{-3}$.  Optical depths have the effect of lowering the critical density since the escape probability multiplies A.   Below the critical density the emissivity varies as the square of the density, and above $n_{crit}$, linearly.  Table 1 (based on Baldwin et al. 1996) gives a list of the more prominent ultraviolet emission lines of quasars along with an indication of their formation mechanism and critical density.

Figure 5 shows the results of changing the density of our standard cloud while keeping the ionization parameter constant. The temperature grows hotter as $n$ increases because denser gas does not cool very efficiently. This is because many of the strong coolants are suppressed above their $n_{crit}$. At the highest densities the cloud has significant free-free opacity at IR wavelengths, and free-free absorption becomes a major heating agent, even with our IR break at 0.912 microns. Ly$\alpha$ becomes weaker relative to the continuum at high densities, as the line becomes *thermalized*, i.e. approaches the black body limit for its emission. Very dense clouds are sources of continuum, not line emission. The lower panel of Figure 5 shows several line ratios that have been used as density indicators. Basically, as all lines become thermalized, their relative intensities approach 1:1 ratios, as each emits near the black body limit for its temperature and wavelength. Note the great strength of Al III] and SiIII] relative to C III] at high densities. Baldwin et al. (1996) argue that these lines are density indicators. The Ly$\alpha$/H$\beta$ ratio is also small at high densities.

## Table 1 Prominent lines in quasars

| Ion | $\lambda$(A) | Components | Formed by | log $n_{crit}$ |
|---|---|---|---|---|
| C III | 977 | 977.03 | rec, coll, pmp | 16.2 |
| N III | 991 | 990.98 | rec, coll, pmp | 15.9 |
| Ly$\beta$ | 1026 | 1025.72 | rec | 15.4 |
| O VI | 1034 | 1031.95, 1037.63 | coll | 15.7 |
| Ly$\alpha$ | 1216 | 1215.67 | rec, coll | 16.9 |
| NV | 1240 | 1238.81, 1242.8 | coll | 15.5 |
| Si II | 1263 | 1260.42, 1264.73 | ?? | 16.4 |
| O I | 1303 | 1302.17, 1304.87, 1306.04 | pmp, coll | 17.0 |
| Si II | 1307 | 1304.37, 1309.28 | ?? | 15.9 |
| C II | 1335 | 1334.53, 1335.66, 1335.71 | coll | 15.7 |
| Si IV | 1397 | 1393.76, 1402.77 | coll | 15.6 |
| O IV] | 1402 | 1397.21, 1399.78, 1404.79, 1407.39 | coll | 11.0 |
| N IV] | 1486 | 1486.5 | coll | 10.2 |
| C IV | 1549 | 1548.20, 1550.77 | coll | 15.3 |
| He II | 1640 | 1640.72 | rec | 16.1 |
| O III] | 1665 | 1660.80, 1666.14 | coll | 10.5 |
| N III] | 1750 | 1748.65, 1752.16, 1754.00 | coll | 10.3 |
| Fe II | UV191 | 1786.7 | ?? | 16.4 |
| Si II | 1814 | 1808.00, 1816.92, 1817.45 | coll | 13.4 |
| Al III | 1857 | 1854.72, 1862.78 | coll | 15.4 |
| Si III] | 1892 | 1892.03 | coll | 11.0 |
| C III] | 1909 | 1908.73 | coll | 9.5 |
| Fe III | UV 34 | 1895.46, 1914.06, 1926.30 | ?? | 16: |
| Fe III | UV 48 | 2061.55, 2068.24, 2078.99 | ?? | 16: |
| N II] | 2142 | 2139.7, 2143.5 | coll | 10.0 |
| Mg II | 2798 | 2795.53, 2802.70 | coll | 15.0 |

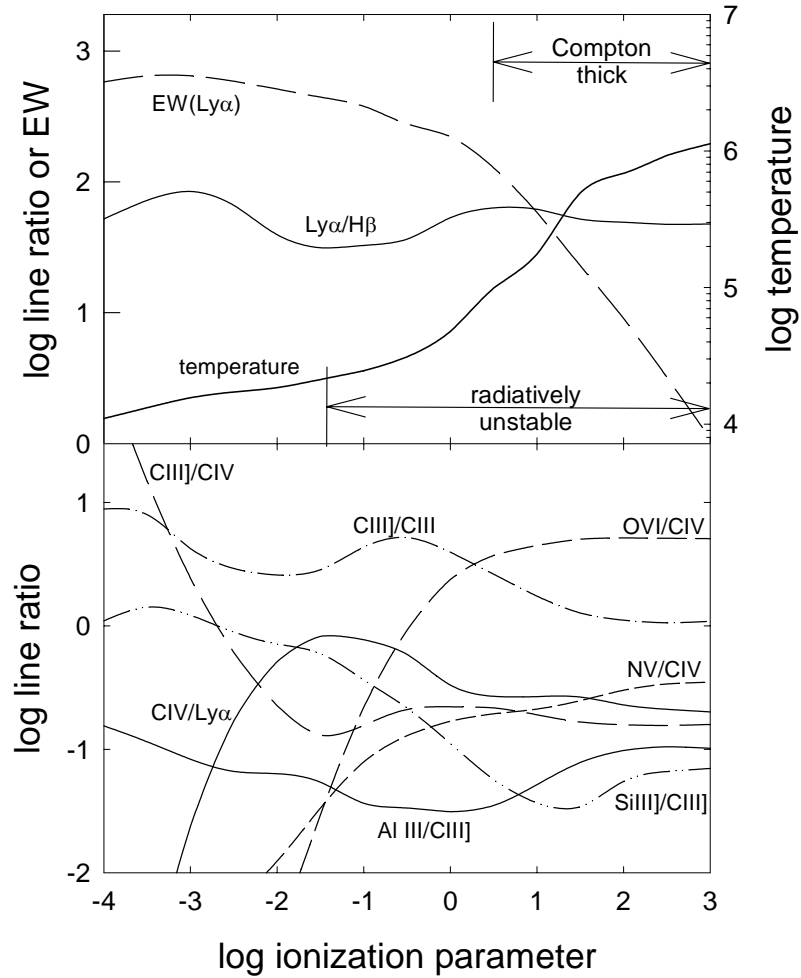

Figure 6 Spectrum as a function of the ionization parameter. Values of the ionization parameter where the cloud is optically thick to electron scattering are indicated as "Compton thick", and where line radiation pressure exceeds gas pressure by "radiatively unstable".

### 3.6 Flux of ionizing photons or ionization parameter

Figure 6 shows the effects of changing the ionization parameter. In these calculations the column density was not held fixed, but the calculation was stopped at the point where the gas temperature fell below 4000 K. The temperature at the illuminated face of the cloud is shown in the upper panel of Figure 6. Both the $O^{++}$-weighted mean T (plotted) and ionization goes up with increasing U.

The ionized column density increases with $U$ (equation 6), and eventually the column density is large enough for the cloud to become optically thick to electron scattering. The figure shows the point where $\tau_e > 1$ and also where the internally generated line radiation pressure exceeds the gas pressure. These clouds would be unstable to disruption by radiation pressure if they are externally supported, such as would happen with a hot intercloud medium (Elitzur and Ferland 1986).

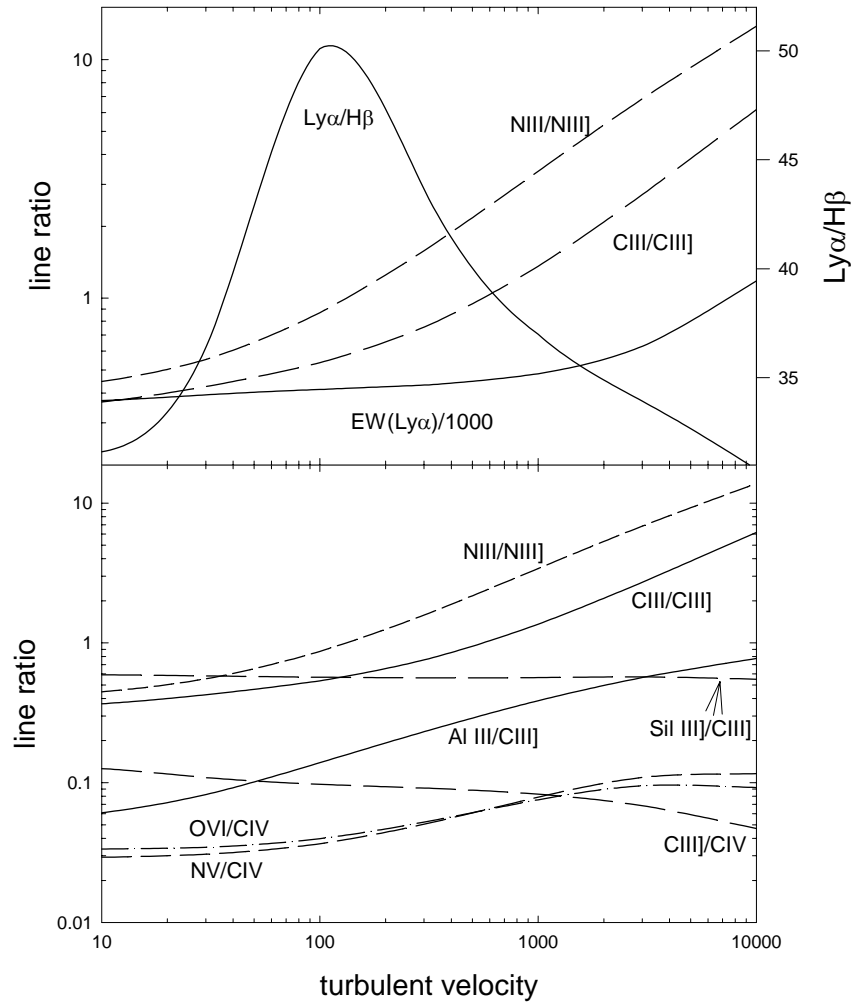

Figure 7  The effects of microturbulent velocity.

For low values of the ionization parameter the level of ionization of the gas, as evidenced by ratios of lines from different ions of the same element, goes up with *U.* This prompted the early statements that the spectrum depended strongly on *U.* Figure 6 shows that the CIII]/CIV ratio is indeed strongly sensitive to *U* for *U* < $10^{-1.5}$. Actually, for high values of *U* the column density of the ionized part of the cloud is so large that individual elements form their own Strömgren shells.  In this limit the line ratios are insensitive to *U* and become more sensitive to the continuum shape.

There has been almost no exploration of these very large U solutions.  The fact that large U clouds can be Compton thick brings in electron scattering as a additional line-broadening mechanism.    This could help explain a long-standing puzzle – the profiles of broad emission lines remain smooth out into the line wings.  If lines from each cloud are only thermally broadened then this smoothness requires an impossibly large number of clouds (Arav et al. 1998). A component of electron broadening would make each cloud's intrinsic line spectrum exceptionally fuzzy and so help this dilemma.

### 3.7 Turbulent velocity

From the beginning, analyses of clouds have assumed that each cloud has only thermal motions, and that the observed very broad line profiles are due to macroscopic motions of the clouds. There has been very little exploration of this assumption.

Figure 7 shows the effects of adding a component of turbulence to our standard cloud. As the line width increases there are two main effects. The first is to decrease line optical depths and so allow resonance lines to escape more freely. They then cool the gas more efficiently. The second is to increase the efficiency of line pumping by the incident continuum. The gas temperature (not plotted) tends to go down, as trapped resonance lines become more efficient coolants at large values of the turbulence. The greatest single effect is to increase the intensities of far UV lines (those shortward of Ly$\alpha$) by adding a pumped component to the (small) component due to collisional excitation.

## 4 Putting it all together -  the curse of the free parameter

The previous discussion should have been more than a little depressing – we have less than a dozen lines whose intensities can be measured in a typical spectrum, and over a dozen pages of free parameters! The miracle is that it is still hard to fit the spectrum.

An approach to understanding the spectrum is suggested by the series of plots shown so far. We have a finite number of free parameters, with the density and flux/ionization parameter having the most powerful effects on the spectrum, and the composition and shape of the ionizing continuum being the most astrophysically interesting. The next step is to isolate which pairs of lines are most sensitive to which parameters – are there line pairs that tell us the density? Baldwin et al. (1996) argue that Al III/C III] is one pair. Others in this volume argue for Si III]/C III], which seems well correlated with Eigenvector 1. Lines of nitrogen relative to C and O lines can tell us the abundances if we think we understand the chemical enrichment history of the environment (Hamann, this volume). The HeII spectrum tells us about the incident continuum at unobservable wavelengths.

Quantitative spectroscopy of emission line sources was identified as a major area of research very early in the history of computational astrophysics (Hjellming 1966, Rubin 1968, Williams 1967, Davidson 1972; MacAlpine 1971). Any one of the figures shown in this paper would have required a large investment of computer time at a major research university in the early 1970's, and it probably would not have been possible for an investigator to do them all. This prevented a complete reconnaissance of parameter space. There are still regimes of parameter space that have not been well explored, as described in previous sections above.

The biggest question for the original workers was why the spectra of quasars are so similar – they do not display the range of line intensities shown by planetary nebulae, for instance. This was taken as an indication that the cloud parameters, especially $U$ and $n$, were always the same. Much of the inital work went into efforts to identify the agents responsible for fine-tuning the parameters. The hot – cold model proposed by Krolik, McKee, & Tarter (1981) was one way to select parameters, but the postulated hot phase has so much opacity that little radiation would escape (Mathews & Ferland 1987). The even softer continuum measured by Zheng et al. (1997) makes this even worse. It is very hard to support the existence of a pervasive hot phase.

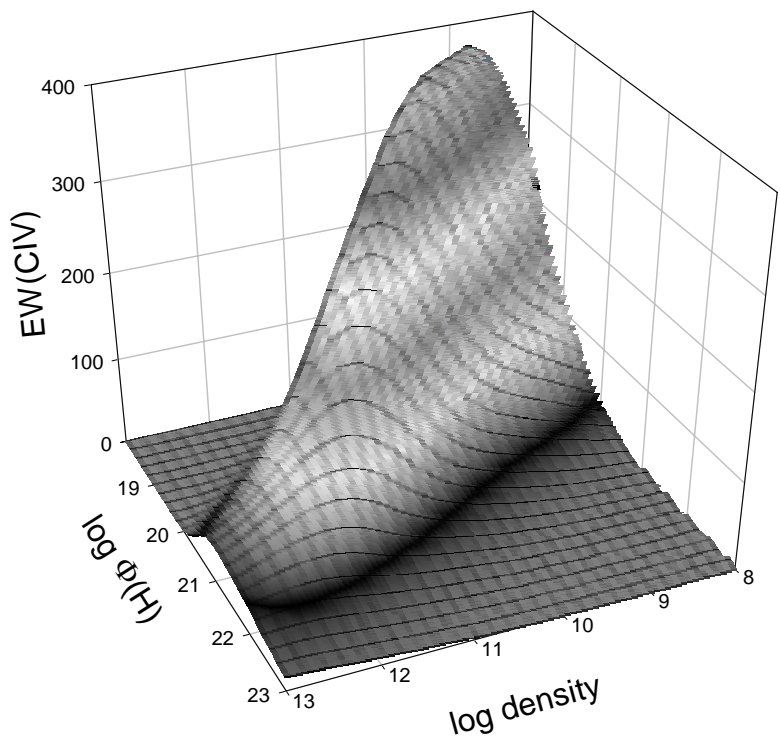

Figure 8  The predicted equivalent width of CIV 1549.

With the power of modern workstations it is now possible to explore far more of parameter space. Figure 8 gives an example, based on Baldwin et al. (1995). It shows the equivalent width of CIV 1549 as a function of the density and flux of ionizing photons. This illustrates the fact that CIV is produced effectively by only a narrow range of parameters, as discussed extensively in the chapter by Kirk Korista in this book. This lead to the LOC approach outlined by Korista, which uses selection effects alone, with no preferred parameters, to fit the observed spectrum. If it turns out that this picture is correct then parameters of individual clouds can be ignored and we can use the clouds to probe more fundamental properties of quasars such as metallicity and the effects of the SED.

## 5 Outstanding problems

There have been great advances in numerical simulations of quasar clouds, largely the result of the power of modern computers. But there are still major questions left unanswered. These are:

♦ *The Lyα/Hβ ratio has not really been solved.* Baldwin (1977) discovered that this ratio is under 10, but all of the plots shown in this paper predict between 30 and 100 (see also Netzer et al. 1995). Several solutions are possible – the first is that the spectrum has been reddened. Another is that the radiative transfer in the current generation of plasma simulation codes is not good enough. Another is suggested by the Kwan and Kro-Krolik (1980) work, which *did* reproduce the observed ratio but with an ionizing continuum that is far harder than is found in AGN – perhaps the actual continuum striking clouds is indeed very hard. Finally, low values of Lyα/Hβ can be produced by LOC integrations that extend to very high densities (Baldwin 1997).

♦ *The BLR is highly stratified.* This is shown by reverberation studies (see the review by Horne in this book). Clearly the type of approach taken here is too simplistic – we must be thinking about the global environment and a range of cloud properties. Rees, Netzer, and Ferland (1989) assumed that the clouds were controlled by an external pressure, which they approximated by a simple power law. Kaspi and Netzer discuss this type of model in a chapter in this book. Another approach is the LOC described by Korista in this book. Here clouds have a very broad range of properties and selection effects introduced by the atomic physics controls the observed spectrum.

♦ *High and low ionization lines do not have the same redshift* (Gaskell 1982; Wilkes 1986; Espey et al. 1989). This is telling us something about the BLR velocity field in a stratified environment. There is not any one model for the origin of the shifts that is accepted, nor any that really works.

♦ *Why are emission line profiles so smooth?* If individual clouds have thermal line widths and the observed line widths are due to bulk motion of a large numbers of clouds, the profiles should break up into individual components far out in the wings were Gaussian statistics say that few clouds should contribute to the profile. This does not happen, leading to the conclusion that an impossibly large number of clouds are involved (Capriotti, Foltz, and Byard 1981; Arav et al. 1998). This could be explained with either electron scattering by relatively cool gas (T < $10^6$ K) or microturbulence.

♦ *Why equivalent width?* Current models of the Baldwin effect assume that it is driven by changes in the continuum shape, perhaps combined with metallicity. If this were all that is going on we would expect *line ratios* to be better correlated with luminosity, and so have less scatter, than a line relative to the continuum. The line equivalent width brings in an additional parameter, the covering factor (eqn 7), which will introduce additional noise unless $\Omega/4\pi$ is always the same. Additionally the continuum will be beamed differently than that of the lines. We have looked for emission-line ratio correlations with luminosity but found none better than the equivalent width correlations. This is unexpected from the nebular physics.

I thank Jack Baldwin and Kirk Korista for their help and comments. NSF and NASA support research in Nebular Astrophysics at the University of Kentucky.